 \documentclass[authoryear,preprint,review,12pt]{elsarticle}

\usepackage{epsfig}

\usepackage{amssymb}
\usepackage{lipsum}

\usepackage{dcolumn}
\usepackage{bm}
\usepackage{subfigure}
\usepackage{siunitx} 
\usepackage{xcolor}
\usepackage{amsmath}

\usepackage[T1]{fontenc}
\usepackage[utf8]{inputenc}

% ...

\journal{Physics Letters B}

\begin{document}

\begin{frontmatter}

\title{New insights on fission of $^{235}$U induced by high energy neutrons from a new measurement at CERN n\_TOF} 

\author[1,2]{A.~Manna\corref{cor1}}%
\ead{alice.manna@bo.infn.it} \affiliation[1]{organization={Istituto Nazionale di Fisica Nucleare, Sezione di Bologna, Italy}} \affiliation[2]{organization={Dipartimento di Fisica e Astronomia, Universit\`a di Bologna, Italy}}
\author[3]{E.~Pirovano\corref{cor1}} %
\ead{elisa.pirovano@ptb.de} \affiliation[3]{organization={Physikalisch-Technische Bundesanstalt (PTB), Bundesallee 100, 38116 Braunschweig, Germany}} %
\cortext[cor1]{Corresponding author}
\author[4,1]{P.~Console Camprini} \affiliation[4]{organization={Agenzia nazionale per le nuove tecnologie (ENEA), Bologna, Italy}}%
\author[5]{L.~Cosentino} \affiliation[5]{organization={INFN Laboratori Nazionali del Sud, Catania, Italy}} %
\author[6,3]{M.~Dietz} \affiliation[6]{organization={School of Physics and Astronomy, University of Edinburgh, United Kingdom}}  %
\author[3]{Q.~Ducasse}  %
\author[5]{P.~Finocchiaro} %
\author[7]{C.~Le Naour} \affiliation[7]{organization={Institut de Physique Nucl\'{e}aire, CNRS-IN2P3, Univ. Paris-Sud, Universit\'{e} Paris-Saclay, F-91406 Orsay Cedex, France}} %
\author[8]{D.~Mancusi} \affiliation[8]{organization={CEA, Centre de Saclay, DEN/DANS/DM2S/SERMA/LTSD, 91191 Gif-sur-Yvette CEDEX, France}} %
\author[1,2]{C.~Massimi}  %
\author[4,1]{A.~Mengoni} %
\author[3]{R.~Nolte}  %
\author[3]{D.~ Radeck}  %
\author[9,10,7]{L.~Tassan-Got} \affiliation[9]{organization={European Organization for Nuclear Research (CERN), Switzerland}} \affiliation[10]{organization={National Technical University of Athens, Greece}}  %
% ...

\date{\today}

\begin{abstract}

The $^{235}$U(n,f) reaction cross section was measured relative to neutron-proton elastic scattering for the first time in the energy region from 10~MeV to 440~MeV at the CERN n\_TOF facility, 
extending the upper limit of the only previous measurement in the literature by more than 200~MeV.
For neutron energies below 200~MeV, our results agree within one standard deviation with data in literature. 
Above 200~MeV, the comparison of model calculations to our data indicates the need to introduce a transient time in neutron-induced fission to allow the simultaneous description of (n,f) and (p,f) reactions.

\end{abstract}

\begin{keyword}
nuclear reaction, neutron-induced reaction, n\_TOF, high energy, fission process, cross section measurement, uranium-235 
\end{keyword}

\end{frontmatter}

\section{Introduction}
\label{introduction}

The current theoretical representations of the fission process largely rely on phenomenological approaches~\citep{Andreyev_2018}, in spite of the advancements of the nuclear models and the abundance of experimental data. 
In fact, the description of the energy dependence of the fission process always depends on empirical parameters. The height of the fission barrier potential and the density of nuclear states at the saddle points are typically adjusted so to reproduce a large set of experimental data, such as mass distributions of residuals, total kinetic energies of fission fragments, spontaneous fission lifetimes, as well as fission cross section data. 

In the context of the reaction models used in the present work, nucleon-induced reactions from around 100~MeV of projectile energies is described as a two-phase process.
As a consequence of the binary interactions between projectile and individual nucleons, a quick intranuclear cascade takes place at the beginning of the reaction process, which results in the emission of few high-energy nucleons.
This first phase comes to its end when a highly-excited residual nucleus forms and, in the second phase of the reaction process, de-excites over a much longer time span through either nucleon evaporation, fragmentation, emission of intermediate-particles or nuclear fission.
Other models, however, treat intermediate-energy fission as a three-phase process that features a pre-equilibrium intermediate stage in-between the two phases. This description was first suggested by~\cite{Gudima1983} and led to the cascade-exciton model. 

For intermediate and high excitation energies (corresponding to neutron/proton energy from $E_{\rm n,p}\!>$\,100~MeV), the intrinsic and collective excitations of the nuclear constituents influence the fission probability. Under these conditions, the dynamic evolution of the highly-excited residual nucleus is the result of a competition between fission and other reaction channels. This situation may lead to a fission hindrance which turns into a fission decay rate smaller than predicted, in some simplified descriptions~\citep{Gavron1986}. 
A possible way to take into account the time-dependent dynamics is to introduce the transient time, i.e. the time required by the fissioning system to attain a quasi-stationary fission decay rate~\citep{Grange1980}. During this time, fission is suppressed, while other decay channels, mostly neutron emission, are open. This neutron emission will cool down the fissioning nucleus and modify its fission barrier, further reducing the fission probability. Although there is no consensus on the necessity to include transient time in the fission dynamics (see i.e. ref.~\cite{Tishchenko2005}), this effect is a possible explanation for the results obtained in several heavy-ion fusion-fission reaction experiments (\cite{Thoennessen1987, Hinde1992, vantHof1996, vantHof1998}), fragmentation of heavy-ion beams (\cite{Ignatyuk1995, Schmitt2007}), and proton-induced fission measurements (\cite{Taieb2003, Jurado2004, RodriguezSanchez2015, RodriguezSanchez2016}). However, it was never observed in neutron-induced fission experiments. In particular, during the last decade, the study of (p,f) cross sections became much more extensive. 
Thanks in particular to experiments in inverse kinematics, the comprehensive isotopic and kinematic characterisation of fission fragments as well as the simultaneous detection of light-charged particles emitted in the fission process has become possible (\cite{Schmidt2000, Schmidt2001, Rodriguez2014, Caamano2013, Farget2013}). 

There is a long-established effort to obtain experimental (n,f) cross section data.
Neutron-induced fission cross sections are typically determined as a ratio with respect to $^{235}$U(n,f), the primary reference for fission reactions. 
For $E_{\rm n}\!>$20~MeV the only available data set covering continuously the energy range up to 200~MeV is that of ~\cite{Lisowski1991}. 
Above 200~MeV, no experimental data exist, and calculations and evaluations rely on theoretical estimates using the $^{235}$U(p,f) reaction as guidance. Even at these energies, the $^{235}$U(n,f) remains the reference reaction for fission studies. 
Therefore, there is a long-standing demand for new $^{235}$U(n,f) cross section measurements extending beyond 200~MeV. 
A new measurement focused on the high energy region 
would allow for a better understanding of the fission process itself, in particular in respect to collective dynamical effects in the reaction.
In addition, considering that fission at high neutron energies of $^{nat}$Pb, $^{209}$Bi, $^{232}$Th, $^{237}$Np and $^{233, 234, 238}$U have been studied relative to the $^{235}$U fission (\cite{Tarrio2023, Paradela2015, Tarrio_2011, Paradela2010}, a re-evaluation of the cross sections for all these cases would be possible with an experimental determination of the $^{235}$U(n,f) cross section above 200~MeV. 
Lastly, neutron-induced fission data at energies above about 50~MeV are required to understand the isospin dependence of the fission cross section. It has been found that the (p,f)/(n,f) cross section ratios changes as a function of the fissility parameter~\citep{Andrey_2002}, however, the role of isospin is still an open question. 
In general, new data in a wider energy range, together with the subsequent further development and refinement of theoretical model descriptions, would allow for important steps forward in basic nuclear science and have considerable relevance in a spectrum of applied fields, including astrophysics, emerging nuclear technologies and space dosimetry.

\section{The \lowercase{n}\_TOF Experimental measurement and analysis}
The measurement was carried out at the CERN white neutron source n\_TOF (neutron time-of-flight), the only source in Europe providing neutrons with energies above 100~MeV~\citep{Rubbia1998}.
At the n\_TOF facility, a proton beam of 20~GeV/c momentum and a narrow pulse width of 7~ns r.m.s. is produced by the CERN Proton Synchrotron (PS) accelerator, with an intensity of $(7-8.5) \times 10^{12}$ proton/pulse, at maximum repetition rate of 0.8~Hz.
Spallation reactions induced by the incident protons on a Pb target, coupled to  water moderators, provide neutron beams to two experimental areas, with energy extending from thermal (25~meV) up to a few GeV.
Energy-dependent cross section measurements, over an extremely wide neutron energy range, can be performed using the TOF method.
The present measurement was performed in the Experimental Area 1 (EAR-1), located at about 185~m from the spallation module. Neutron beam-line elements leading to EAR-1 include a beam-size reduction shielding and collimator, a magnet for charged-particle removal and a second shaping collimator (18~mm diameter). 
The main features of the neutron beam at n\_TOF EAR-1 are a beam size of 2~cm diameter, an intensity of about~$10^{6}$ neutrons/pulse, and high energy resolution $\Delta E / E$ that ranges from $10^{-4}$ at thermal energy to $10^{-2}$ in the hundreds of MeV region~\citep{Guerrero2013}. 

\begin{figure}[hbt]
\includegraphics[width=15cm]{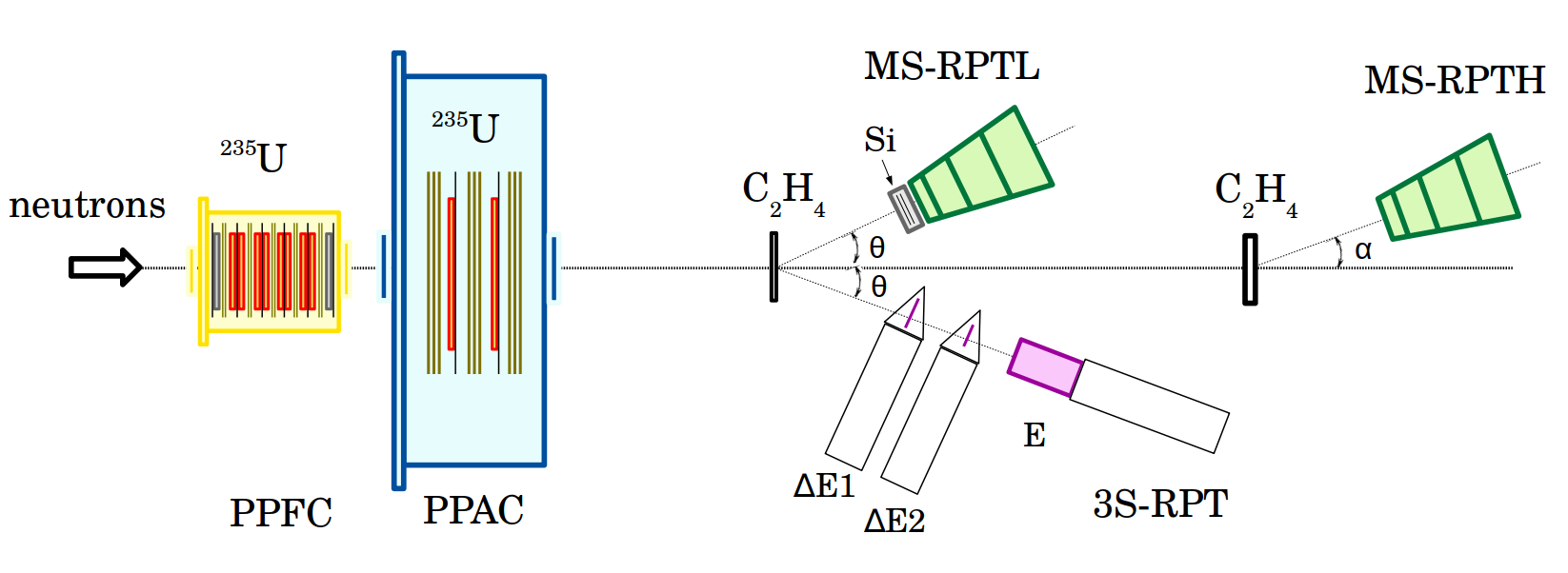}
\caption{\label{fig:setup} Schematic view of the experimental setup, with a parallel-plate fission chamber (PPFC) and a parallel-plate avalanche counter (PPAC), used to detect fragments from fission events. Two multi-stage recoil proton telescopes (MS-RPTL and MS-RPTH at 25$^\circ$ and 20$^\circ$, respectively) and the triple-stage recoil proton telescope (3S-RPT at 25$^\circ$) were adopted to measure the neutron flux impinging in the $^{235}$U samples. }
\end{figure}

The experiment aimed at the determination of the absolute $^{235}$U(n,f) cross section at neutron energies above 10~MeV with a redundant approach, where by convention a cross section measurement is labelled as `absolute' when measured relative to the n-p elastic scattering cross section, the primary standard for neutron cross section measurements~\citep{Carlson2011}. 
The experimental setup was therefore designed to measure simultaneously neutron-induced fission events and proton recoil events (i.e., the incident neutron flux), using five detectors combined to obtained two independent detection systems (Fig. \ref{fig:setup}). 
A detailed discussion of the performances of the used detectors and analysis methods can be found in refs.~\cite{Manna2023} and~\cite{Pirovano2023}; here only the most important aspects will be recalled.

To detect the fission events, a total of ten highly-enriched $^{235}$U samples (two at 92.699(5)\%, eight at 99.933(14)\%) were mounted in two ionization chambers, a Parallel Plate Fission Chamber (PPFC) and a Parallel Plate Avalanche Counters (PPAC).

The PPFC consisted of a stack of eight $^{235}$U samples with a diameter of 42~mm and areal densities ranging between 264\,\unit{\micro g/cm^2} and 338\,\unit{\micro g/cm^2}, measured via alpha-counting from $^{235}$U decay. The samples were grounded and used as cathodes, and placed at a distance of 5~mm from the anodes.
The main advantages of the PPFC detector are the simplicity and the high geometrical efficiency for fission events determination, ideally $2 \pi$. Inefficiencies caused by fission fragments stopped in the finite thickness of the uranium layer and the counting gas, by the fission-fragment anisotropy and the partial transfer of linear momentum were determined with a dedicated Monte Carlo simulation.
The detection efficiency of the PPFC for the fission fragments, averaged over all samples, varies between 90.7\% and 91.7\%. Only data for incident neutron energy up to $E_{\rm n}\!\leq\!145$~MeV were used in the analysis, since the main limitation of the PPFC detector is represented by possible background from neutron-induced reactions on the aluminum backings~\citep{Pirovano2023}. The PPFC confirmed to be a robust instrument in this energy range.

Each PPAC detection unit is composed of two cathodes and one anode. For this experiment, the PPAC chamber consisted of three PPAC detection units and two 80~mm diameter $^{235}$U samples, for which the measured areal density resulted in 263~\unit{\micro g/cm^2} and 282~\unit{\micro g/cm^2}, respectively. 
The PPAC is capable to detect fission fragments produced by neutrons with energies from thermal to a few GeV. The signature for a fission event is the coincidence between the two PPAC detection units on either side of a $^{235}$U sample, generated by the two fission fragments emitted in back-to-back directions. Thanks to the stripped cathodes, the energy-dependent detection efficiency, including the fission-fragment anisotropic emission and the Lorentz boost, could be determined experimentally. The efficiency imposing, the coincidence condition, turns out to be $56\!-\!63\%$ and $48\!-\!56\%$, respectively for the two uranium samples~\citep{Manna2023}.

The n-p elastic scattering reaction, considered the primary reference for fast neutrons measurements~\citep{Carlson2009}, was employed for the determination of the neutron flux. To cover the incident neutron energy range of several orders of magnitude, two polyethylene (C$_2$H$_4$, PE) sheets were simultaneously irradiated, with the recoil protons detected by three Recoil Proton Telescopes (RPTs) placed at 20$^\circ$ and 25$^\circ$ relative to the beam direction.
The samples were characterized at PTB (Physikalisch-Technische Bundesanstalt), where the measurements of the density, thickness, elemental composition and the stoichiometric relation of hydrogen to carbon atoms, were performed~\citep{Pirovano2023}. The RPTs, which were specifically developed and characterised for this experiment, followed two design concepts focused on the neutron energy range covered: one from 28~MeV to 145~MeV (triple-stage recoil proton telescope, 3S-RPT), the other two for energies up to 440~MeV (multi-stage recoil proton telescopes, MS-RPTL and MS-RPTH). 
They were assembled using mainly plastic scintillators (BC-408 and EJ-204) to benefit from their fast response of a few nanoseconds rise-time and 600~ps time resolution, characteristics required to detect particles in the high-energy region, where kinematics is compressed.
The analysis of the RPTs responses was based on selecting coincident signals between consecutive sub-detectors, which allowed to discriminate events generated at the sample position, remove background events and achieve the required particle identification. \\
The main reactions originating from the interaction of neutrons on C$_2$H$_4$ are $^{1}$H(n,p), $^{12}$C(n,d), $^{12}$C(n,t), $^{12}$C(n,$\alpha$), and $^{12}$C(n,p), thus producing different charged particles in the exit channel. Therefore, their ability to discriminate protons from other light charged particles with the $\Delta E$-$E$ technique was the basis for the choice of RPTs as flux detectors. Even with the particle discrimination capability of the RPTs, an unavoidable background contribution from the $^{12}$C(n,p) reaction events remains to be evaluated. To this end, an additional measurement using graphite samples with an areal density similar to that of carbon in C$_2$H$_4$, was performed.
The number of proton counts was obtained as a function of the neutron energy by selecting the protons via their characteristic $\Delta E-E$ signature. The contribution from $^{12}$C(n,p) reactions was obtained and subtracted from the measurement with the graphite target, after normalising to the number of incoming neutrons and the respective number of atoms of carbon in the two targets. 

The experimental fission cross section, $\sigma_{\rm f}$, was determined from the relation
\begin{equation}
    \sigma_{\rm f} (E_{\rm n}) = \frac{C_{\rm f}(E_{\rm n})}{C_{\rm p}(E_{\rm n})} \; \frac {\varepsilon_{\rm p} (E_{\rm n}) \;  n_{\rm H} \;  \Omega_{\rm p}}{ \varepsilon_{\rm f}(E_{\rm n}) \;  n_{\rm F}} \; \frac {{\rm d} \sigma_{\rm n,p}(E_{\rm n})} {{\rm d} \Omega_{\rm p}} (\theta_{\rm p}).
\end{equation}

Here, $C_{\rm f}$ and $C_{\rm p}$ are the number of detected fission and recoil proton events per bunch produced by neutrons with energy $E_{\rm n}$, respectively. $\varepsilon_{\rm f}$ and $\varepsilon_{\rm p}$ are the detection efficiencies for fission fragments and recoil protons, respectively, $\Omega_{\rm p}$ is the solid angle covered by the RPT, $n_{\rm H}$ and $n_{\rm F}$ are the number of hydrogen and fissile atoms per unit area, respectively. ${\rm d} \sigma_{\rm n,p}/ {\rm d} \Omega_{\rm p}$ is the differential cross section for the emission of a recoil proton at the laboratory angle $\theta_{\rm p}$, for which the VL40 PWA energy-dependent phase-shift solution by~\cite{Arndt1991} and collaborators, the reference recommended by the International Nuclear Data Commitee~\citep{Carlson1997}, was adopted. 

\begin{figure}[hbt]
{\includegraphics[width=15cm]{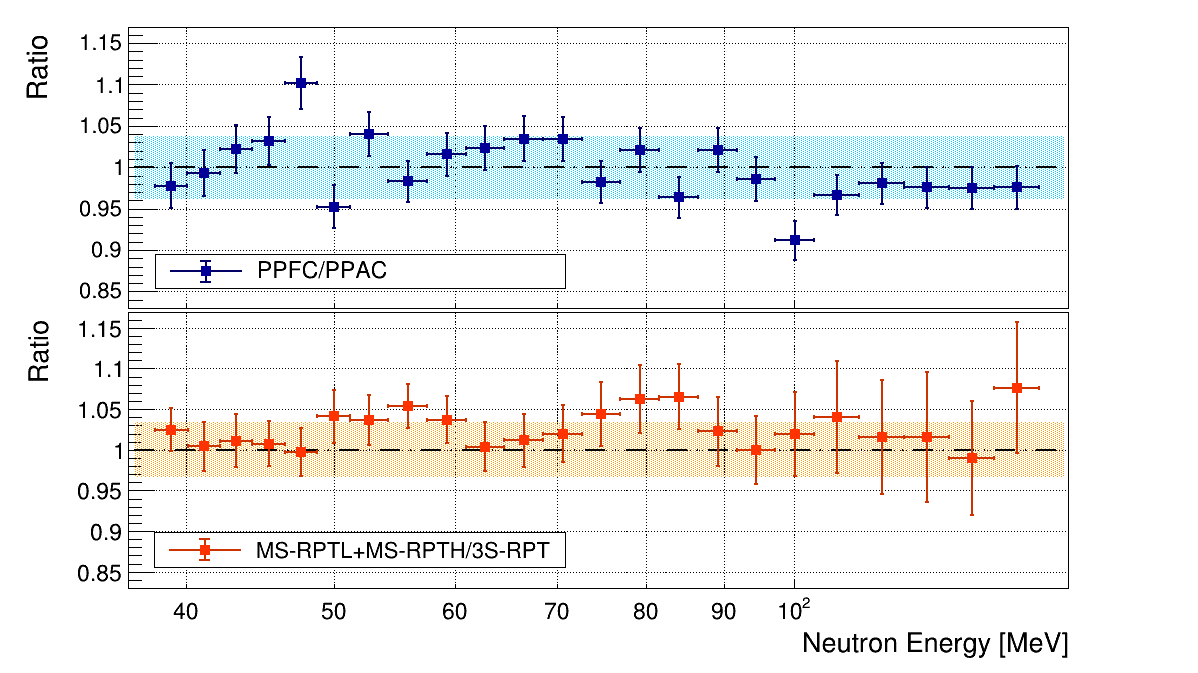}}
\caption{Yield ratio obtained with different detection systems: PPAC vs PPFC (top panel), 3S-RPT vs MS-RPTL+MS-RPTH (bottom panel). The coloured bands represent the combined systematic uncertainties: 3.4\% and 3.8\% in the fission yield and in the neutron flux, respectively.}
\label{Fig:ratio}
\end{figure}

The analysis of the two detection systems -- PPFC with 3S-RPT for $E_{\rm n}=$~28--145~MeV and PPAC with MS-RPTL+MS/RPTH for $E_{\rm n}=$~10--25~MeV and 39--440~MeV -- was carried out independently and the results were compared and combined after applying all the corrections required by each of the two apparatus. The two detection systems provided the shape and the absolute value of the $^{235}$U(n,f) cross section, without the need of any re-normalization. The energy region in which the data of the two detection systems overlap determines the consistency and reliability of the results. 
Fig.~\ref{Fig:ratio} shows the ratio between the results of the two independent systems as a function of the incident-neutron energy, both for fission counts (top panel) as well as for the neutron flux (bottom panel). The plots indicate an excellent agreement between PPFC and PPAC yields, and between the neutron flux obtained via 3S-RPT and the weighted average between MS-RPTL and MS-RPTH,
calculated separately with the differential cross section at the appropriate angle.
No normalisation was applied. Both ratios are consistent within the combined statistical (error-bars in Fig.~\ref{Fig:ratio}) and systematic (colored bands in Fig.~\ref{Fig:ratio}) uncertainties. 
A detailed budget of the evaluated systematic uncertainties is provided in refs.~\cite{Manna2023} and~\cite{Pirovano2023}.

\section{Discussion}
\begin{figure*}[h!]
\subfigure{\includegraphics[width=8cm]{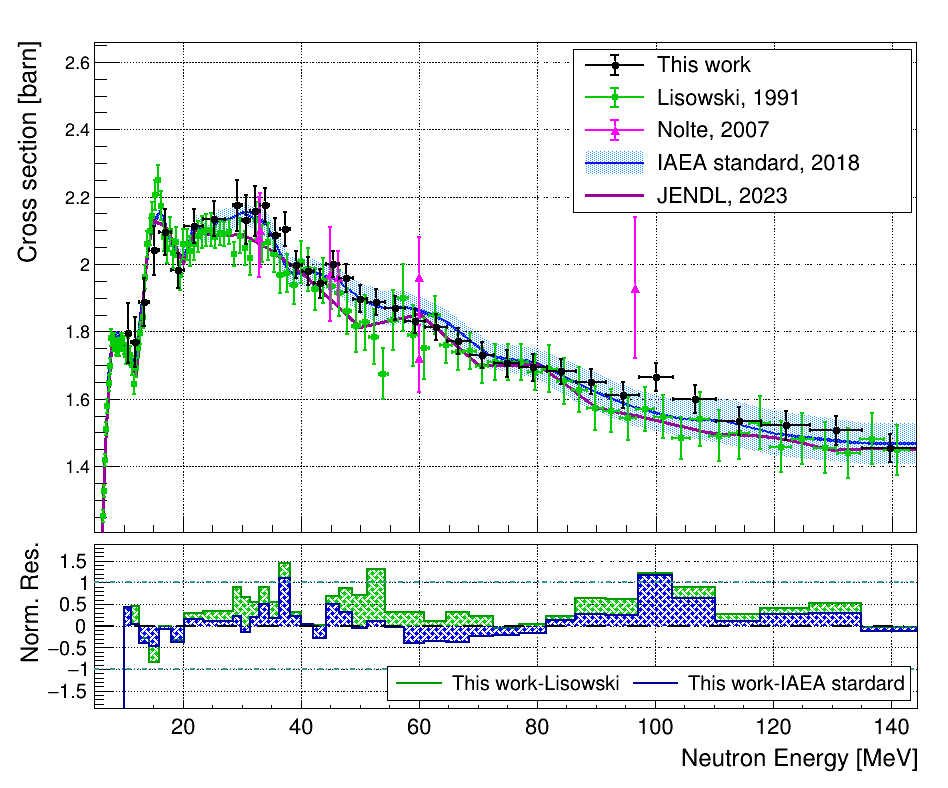}\label{fig:cross_section_LE}}
\subfigure{\includegraphics[width=8cm]{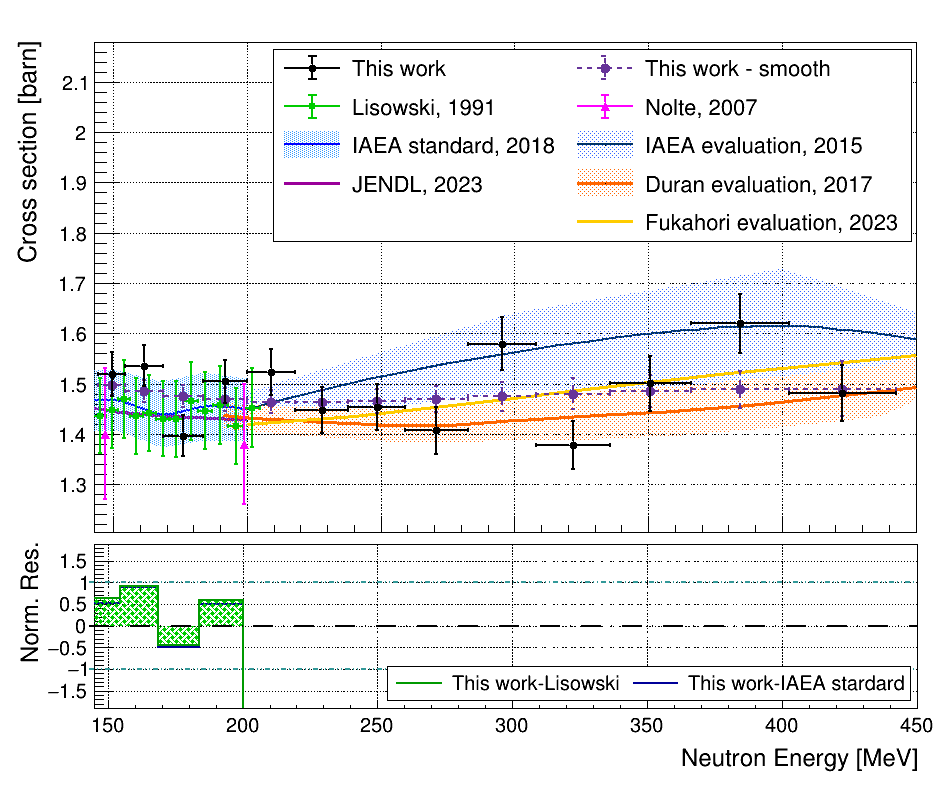}\label{fig:cross_section_HE}}
\caption{(a) Cross section obtained with the weighted average of the two detection systems. The \cite{Lisowski1991} and \cite{Nolte2007} data together with JENDL-5~\citep{Iwamoto2023} and IAEA evaluations~\citep{Carlson2018}, which define the standard cross section up to 200~MeV, are reported. The error bars represent the total uncertainties given by the sum in quadrature of the statistic and the systematic components. 
(b) Cross section obtained with PPAC and MS-RPTL+MS-RPTH, together with the data sets in literature and IAEA~\citep{Marcinkevicius2015}, \cite{Duran2017}, JENDL~\citep{Iwamoto2023} and \cite{Fukahori2023} evaluations.}
\label{Fig:cross_section} 
\end{figure*}

We present the final results of the neutron-induced fission cross section at energies covered by both detection systems in Fig.~\ref{fig:cross_section_LE}.
The cross section expectation value $\bar{\sigma}_{\rm f}$ and its uncertainty were calculated from the posterior probability distribution accounting for the correlated uncertainties induced by the use of the same C$_2$H$_4$ targets for the measurements with the 3S-RPT and the MS-RPTL, and by the same beam-transmission factor through the fission detectors (placed upstream of the PE sample)~\citep{Neudecker2012, PirovanoManna2023}.
Fig.~\ref{fig:cross_section_HE} shows the cross section for neutron energy above 145~MeV, where we rely only on the MS-RPTL+MS-RPTH and PPAC detection system. 

In Fig.~\ref{Fig:cross_section} our results are compared to the literature experimental data sets  (\cite{Lisowski1991}, \cite{Nolte2007}), as well as to the IAEA [\cite{Carlson2018, Marcinkevicius2015}], JENDL [\cite{Iwamoto2023}], \cite{Duran2017} and \cite{Fukahori2023} evaluations. 
The bottom panels show the normalised residuals between our cross sections data and both Lisowski, and the IAEA standards evaluation, respectively. The agreement is within 1 standard deviation. 

\begin{figure*}[bt]
\subfigure[]{\includegraphics[width=8.1cm]{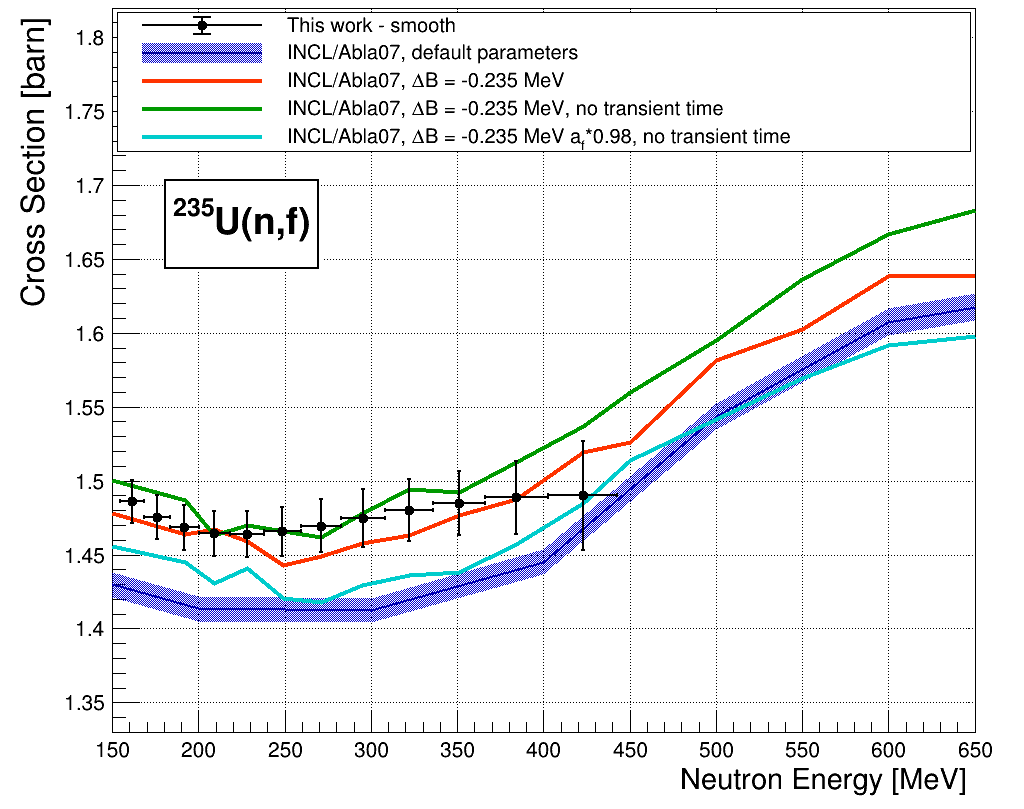} \label{fig:xs_n}}
\subfigure[]{\includegraphics[width=8.1cm]{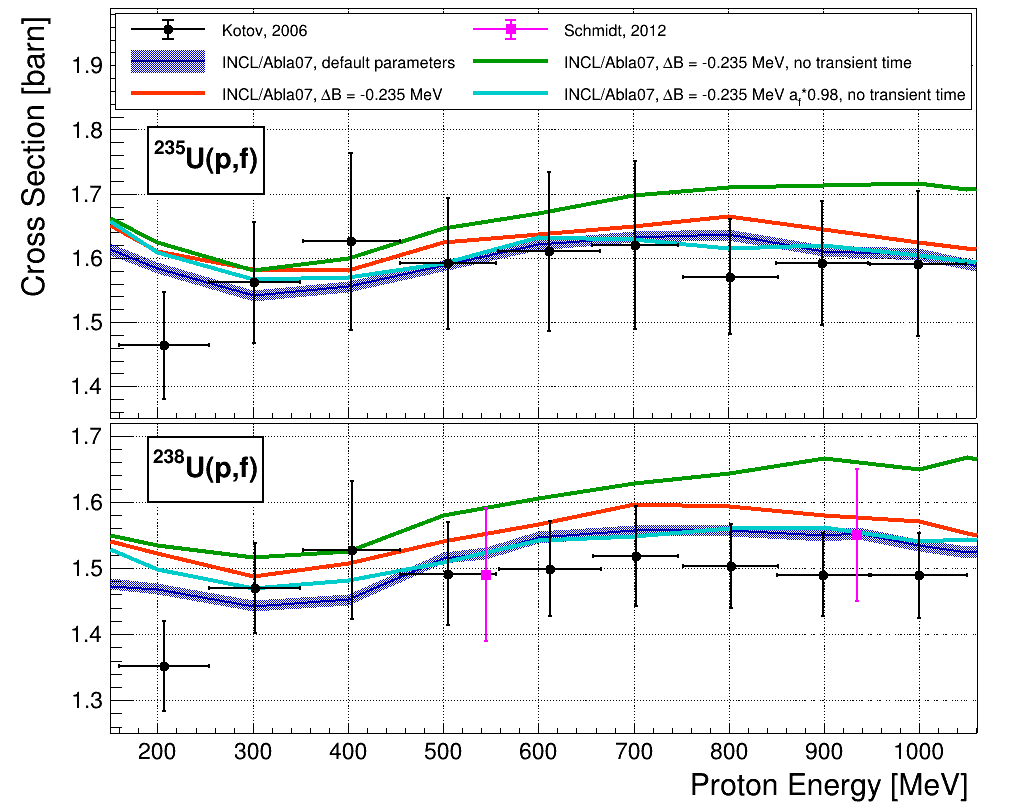}\label{fig:xs_p}}
\caption{(a) The smoothed result of this work is reported in black dots with the related fit uncertainty (see text). The cross section is compared with four calculations based on INCL++~\citep{INCL2020} coupled with ABLA07~\citep{Kelic2008}, using: i) default values for fission barrier height and level density parameter at the saddle point, $a_f$, for all spallation remnants; ii) changing the parameters (the barrier height $\Delta B_f$=-0.235~MeV and 0.995$a_f$) and iii) two calculation without including the effect of the transient time with 0.995$a_f$ and 0.98$a_f$, respectively. (b) Results of the same calculations for (p,f) cross section compared to \cite{Kotov2006} and \cite{Schmidt2012} data.}
\label{fig:cross_section_2}
\end{figure*}

Above 145~MeV, the results suffer from larger statistical fluctuations, therefore, to guide the eye, a data-smoothing process was applied to the fission yield and to the neutron flux, separately. This operation is based on the assumption that no structure is expected in neither of the two. The resulting cross section is shown in Fig.~\ref{fig:xs_n}. 
The smoothing process consists on the fit of the data with a polynomial from the 3$^{rd}$ to the 7$^{th}$ order, as a function of log($E_n$). The shape of the curves obtained with the three higher-degrees polynomial fit does not change, establishing the convergence of the smoothing process. The reduced $\chi^2$ obtained, from both the detectors separately, was 0.95, 0.98 and 1, for the 5$^{\rm th}$, 6$^{\rm th}$ and 7$^{\rm th}$ degrees respectively, confirming a statistical dispersion in agreement with statistical uncertainties and the shape of the fit.

This work resulted in the first absolute experimental cross section of the neutron-induced fission on $^{235}$U above 200~MeV. With this, a new comparison can be performed between theoretical models and data.
In Fig.~\ref{fig:cross_section_2}, (n,f) and (p,f) measurements are compared with four different calculations performed with version 6.29 of INCL++~\citep{Mancusi2014} coupled with ABLA07~\citep{Kelic2008}.
The approach used in these calculations is explained in~\cite{LoMeo2015}, wherein INCL++ describes the intranuclear cascade leaving an excited pre-fragment or compound nucleus, and the subsequent de-excitation is handled by the ABLA07 code. 
These calculations are expected to be appropriate above 150 MeV but, so far, they could have been assessed on the neutron induced fission on $^{235}$U only in the limited energy range covered by the experimental data previously available (up to 200 MeV).
As it can be seen from the figure, the calculations with default values for fission barrier height and level density parameters of fission remnants clearly underestimate the measured $^{235}$U(n,f) cross section. By changing these two parameters, it is possible to reproduce the absolute value of the measured cross section. 
Reducing the fission barrier heights of all spallation remnants by the same amount, $\Delta B_f = -0.235 $ MeV is a rough approximation, but is consistent with the uncertainties of experimentally determined barrier heights of actinides, usually of the order of 0.2 - 0.3 MeV (see Table XXXII of ref.~\cite{BL80}); a small reduction of the asymptotic level density parameters at the saddle points, $a_f$, partially counterbalances the effect of the reduction of barrier heights mainly at higher excitation energies.
Moreover, in our calculations the average remnant excitation energy shows a fast increase with incident energy in the range from 200 MeV to 1 GeV: this might lead to an energy-dependent reduction of barrier heights and to a consequent increase of fission widths of spallation remnants ( see, {\it e. g.}, ref.~\cite{DS18}); however, introducing the formalism of energy dependent fission barriers in the ABLA07 code goes beyond the scope of the present work.\\
With the same set of parameters, we expect to reproduce the $^{235}$U(p,f) cross section, for which \cite{Kotov2006}'s work is the main reference, and, given their proximity, also the $^{238}$U(p,f) cross section measured by~\cite{Kotov2006} and~\cite{Schmidt2012} in direct and inverse kinematics, respectively. 
An additional parameter included in the calculation is the transient time~\citep{Grange1980}, which encompasses the dynamical delay of the fission process due to dissipation. The transient time becomes relevant when the initial projectile energy, therefore the excitation energy, reaches several hundreds MeV. 
In ABLA07, the effect of dissipation is modelled using a Fokker-Planck equation for the deformation parameter, with a parabolic approximation of the nuclear potential. The effect of the ground-state nuclear deformation is taken into account by suitably chosen initial conditions (\cite{Jurado2005, Jurado_2005}).
In the calculations done with the default parameters the effect of the transient time is included also in neutron-induced fission, although it has never been observed. 

While proton-induced fission data sets can also be reproduced by switching off the transient time effect and adjusting the level density, the only way to achieve simultaneous consistency of $^{235}$U(n,f) and $^{235, 238}$U(p,f) cross section data, is to consider the effect of the transient time and a slight change in the level density parameter with respect to the default value, as visible in Fig.~\ref{fig:cross_section_2}. 
The curves obtained with this approach represent, in fact, the best-fit between the calculations and data from all the experimental data sets.
Our data provide a first indication for this effect to take place in neutron-induced fission reaction.

\section{Conclusion and outlook}
In conclusion, the absolute $^{235}$U(n,f) cross section was measured at the n\_TOF facility at CERN for the first time up to 440~MeV of neutron energy,
 increasing the previous experimental energy limit by 240 MeV.
Two fission chambers and three neutron flux detectors were developed and optimized for this experiment.
The redundant approach, used in designing the experimental setup, together with the agreement of the results from the independent analysis of the detection systems, ensure the reliability of the obtained data. The extracted cross section, in the region 10-200~MeV neutron energy extends the scarce amount of data available. 
At higher neutron energies, from 200 to 440~MeV, the results of this pioneering measurement are the first available experimental data. 
Our data provided the opportunity to refine the theoretical model description of the neutron-induced reaction, constraining the behaviour of the fission cross section at high excitation energy. An indication of the time-dependent diffusion effect (transient time) in the neutron channel has been observed for the first time. \\
The physics outcomes of our results, open the possibility to investigate the transient time effect in the neutron-induced fission process. Considering that the impact of the reaction mechanism is expected to increase with the excitation energy, at n\_TOF even higher energies can be reached as neutrons are available up to the GeV range and fission events have been already detected up to these energies~\citep{Paradela2015, Tarrio2023}.
Increasing further toward high energy the cross section measurement will require the development of a proton detection system able to discriminate the elastic as well as the inelastic n-p scattering reaction processes, expected to play a non-negligible role at energies above 440~MeV. In fact, the upper energy limit of the present experiment is mostly defined by the opening of the inelastic channel in the n-p scattering process.
A brief outlook of future activities at n\_TOF includes a simultaneous measurement of the cross section and fission fragments angular distribution for better constraining the theoretical models at energies beyond the range explored in the present work.



\bibliographystyle{elsarticle-harv} 
\bibliography{u235pap2}


\end{document}